# Performance of Monolayer Graphene Nanomechanical Resonators with Electrical Readout

Changyao Chen[1]†, Sami Rosenblatt[2]†, Kirill I. Bolotin[3], William Kalb[1], Philip Kim[3], Ioannis Kymissis[2], Horst L. Stormer[3,4,5], Tony F. Heinz[2,3], James Hone[1]*

[1]*Department of Mechanical Engineering, Columbia University, New York, 10027, USA*

[2]*Department of Electrical Engineering, Columbia University, New York, 10027, USA*

[3]*Department of Physics, Columbia University, New York, 10027, USA*

[4]*Department of Applied Physics, Columbia University, New York, 10027, USA*

[5]*Bell Labs, Alcatel-Lucent, Murray Hill, New Jersey 07974, USA*

†*These authors contributed equally to this work*

*\*e-mail: jh2228@columbia.edu*

**The enormous stiffness and low density of graphene make it an ideal material for nanoelectromechanical (NEMS) applications. We demonstrate fabrication and electrical readout of monolayer graphene resonators, and test their response to changes in mass and temperature. The devices show resonances in the MHz range. The strong dependence of the resonant frequency on applied gate voltage can be fit to a membrane model, which yields the mass density and built-in strain. Upon removal and addition of mass, we observe changes in both the density and the strain, indicating that adsorbates impart tension to the graphene. Upon cooling, the frequency increases; the shift rate can be used to measure the unusual negative thermal expansion coefficient of graphene. The quality factor increases with decreasing temperature, reaching ~$10^4$ at 5 K. By establishing many of the basic attributes of monolayer graphene resonators, these studies lay the groundwork for applications, including high-sensitivity mass detectors.**

Since its discovery in 2004[1], graphene has attracted attention because of its unusual two dimensional (2D) structure and potential for applications[2-4]. Due to its exceptional mechanical properties[5] and low mass density, graphene is an ideal material for use in nanoelectromechanical systems (NEMS), which are of great interest both for fundamental studies of mechanics at the nanoscale and for a variety of applications, including force[6], position[7] and mass[8] sensing. Recent studies using optical and scanned probe detection have shown that micron-size graphene flakes can act as MHz-range NEMS resonators[9,10]. Electrical readout of these devices is important for integration and attractive for many applications. In addition, characterization of the basic attributes of these devices, including their response to applied voltage, added mass, and changes in temperature, allows detailed modeling of their behavior, which is crucial for rational device design.

Samples are fabricated by first locating monolayer graphene flakes on Si/SiO$_2$ substrates, then patterning metal electrodes and etching away the SiO$_2$ to yield suspended graphene. The ability to choose monolayers in advance provides control of device properties and facilitates electrical readout. The fabrication method also provides control over the lateral dimensions; devices can be either micron-wide sheets (Fig. 1a) or lithographically defined nanoribbons (Fig.1b). Because the etchant diffuses freely under the sheets, the SiO$_2$ is removed at the same rate everywhere under the graphene, so that the distance between the substrate and the suspended sheet is constant (~100 nm) across each device. For the same reason, the portion of each electrode that contacts the graphene is also suspended[11,12], as depicted in Fig. 1c.

Following previous work[13-15], we implemented an all-electrical high-frequency mixing approach (Fig. 1d) to actuation and detection of mechanical resonances. A DC gate voltage $V_g$ applies static tension to the device, an RF gate voltage with amplitude $\delta V_g^f$ at frequency $f$ drives the motion, while a second RF voltage, at a slightly offset frequency $f + \Delta f$, is applied to the source. Because the graphene conductance changes with distance from the gate, motion is detected as a mixed-down current $I^{\Delta f}$ at the difference frequency $\Delta f$. The weaker gate response of the conductivity of multilayer graphene makes the use of monolayers advantageous for this method.

Figure 2a shows $I^{\Delta f}$ as a function of drive frequency for a 3 μm wide, 1.1 μm long monolayer graphene resonator (device 1) at $V_g = -7V$, measured at room temperature. A prominent peak is observed near 65 MHz, which we interpret as the mechanical resonance of the graphene. Smaller peaks, which we assign to the resonance of the under-etched gold electrodes, appear near 20 and 25 MHz. The inset shows the same graphene resonance driven at a lower amplitude and fit to a Lorentzian lineshape. The width of the resonance corresponds to a quality factor Q = 125. This value is typical for the >20 devices studied so far and consistent with previous results[9].

The evolution of the resonances with $V_g$ and device geometry supports the assignment of the two modes given above. The graphene resonance (Fig. 2b) is highly tunable, increasing in frequency away from a minimum near $V_g = 0$ due to the tension induced by the gate voltage (the small observed shift of the resonant frequency minimum away from $V_g = 0$ can be explained by trapped charges in the oxide and on the graphene[13,14]). On the other hand, the resonances of the much stiffer gold beams are, as expected, independent of gate voltage. The contrast between the two resonances serves to emphasize that the combination of high frequency and high tunability with an electrostatic gate is a unique feature of devices with thickness near the atomic scale such as graphene and single-walled carbon nanotubes[14]; it is seen in neither multilayer graphene[9] nor in NEMS resonators fabricated using top-down techniques. For devices with suspended graphene lengths $L$ from ~0.5-2 μm, the graphene resonant frequency scales approximately as ($1/L$), as expected for a thin membrane (see Supplementary Information). The resonant frequency of the gold, on the other hand, scales approximately as ($t/L_{gold}^2$), where $t$ is the electrode thickness, and $L_{gold}$ is the length of the suspended section of the gold electrode (~width of the graphene sheet), as expected for a thick beam. For narrow graphene ribbons, such as in Figure 2c, the gold beam resonance occurs above the measured frequency range, and only the graphene resonance is observed.

The observation of the metal resonances by electromechanical mixing implies that the motion of the metal electrode is transferred to the graphene, since the conductance of the gold is independent of gate voltage. As expected, the coupling between the two resonant modes is strongest when the frequencies are similar. This behavior is manifested in many devices as avoided crossings at intersections between the metal and graphene modes, as indicated by the arrows in Fig. 2a. (and appearing more prominently in other devices, such as shown in Figs. 4 and 5). In addition to being of fundamental interest as a coupled nanoscale-microscale system, these resonances demonstrate that graphene can be used to transduce the motion of larger resonant systems with minimal damping.

Figure 2d shows the evolution of the graphene resonance (for device 1) with drive amplitude. As $\delta V_g$ increases, the peak grows in height, shifts upward in frequency, and changes from a symmetric Lorentzian to an asymmetric shape characteristic of a nonlinear response. Above a drive amplitude of ~44 mV, corresponding to an oscillation amplitude of ~1.1 nm, the resonance develops bistable behavior[14] (see Supplementary Information). The ratio of the amplitude at the onset of nonlinearity to the noise floor gives a dynamic range[16] of approximately 60 dB. Importantly for applications of these devices, the measured peak current values are ~2 orders of magnitude larger than those observed in carbon nanotubes for the same applied power at room temperature[14,17]. This improvement in signal levels is a direct consequence of the ability to fabricate micron-wide devices with higher conductance than that of a one-dimensional nanotube.

**Mechanical modeling**

Modeling the behavior of the graphene resonators is crucial for device design and interpretation of experimental results. We use a continuum model that treats the graphene resonators as membranes with zero bending stiffness[5,18,19,20] (see Supplementary Information). In this model, the resonant frequency is given by

$$f_{res}(V_g) = \frac{1}{2L}\sqrt{\frac{T_0 + T_e(V_g)}{\rho w}} \qquad (1)$$

where $L$ and $w$ are the length and width of the graphene sheet. The two-dimensional mass density $\rho$ represents the sum of the contributions from the graphene and any adsorbates, and therefore in general $\rho \geq \rho_{graphene} = 7.4 \times 10^{-19} g/\mu m^2$. Near $V_g = 0$, the frequency is set by the built-in tension $T_0$ and $\rho$, while at higher $V_g$ the electrostatically-induced tension $T_e$ causes the frequency to increase roughly as $V_g^{2/3}$. The model successfully describes the gate voltage dependence of the resonant frequency in all of the >20 devices studied so far (with lengths from ~0.5-2 µm and widths from 0.2-2 µm); two representative curves are shown in figure 3a. Furthermore, it can be used to deduce both the mass density and the built-in tension in the graphene sheet, which are left as fitting parameters. The solid black points in Fig. 3b show the built-in tension (converted to strain using the measured stiffness of graphene[5] in order to normalize for the geometry) and density (normalized to the density of a graphene sheet) for 11 devices, as derived from the curve fitting. In all cases, a density larger than that of pristine graphene is required to fit the observed data. We attribute this extra mass to e-beam resist (Poly(methyl methacrylate), PMMA) residue from the fabrication process[21,22]. The built-in strain is positive, and of order $10^{-4}$ in all of the as-fabricated devices, which accounts for the observed large resonant frequencies even near $V_g = 0$. Although there is considerable scatter in the data, we observe that the built-in strain is generally larger in devices with more adsorbed residue, which suggests that the PMMA tends to impart a tension to the graphene.

The measured strain can also be used to predict the amplitude at the onset of nonlinearity $a_c$. By modifying an expression previously derived for beams[23] we obtain for a membrane:

$$a_c = 0.56L\sqrt{\frac{\varepsilon}{Q}} \qquad (2)$$

where $\varepsilon$ is the strain of the resonator. For the device studied in Fig. 2d, this expression gives $a_c = 1.5nm$, very close to the measured value.

The validity of the continuum model is further supported by two experiments described below. In the first, a calibrated mass is added to the resonator, while the second uses thermal contraction to change only the tension. In both cases, the model correctly predicts the change in resonator mass.

**Response to changes in mass**

Ohmic heating of suspended graphene in vacuum has previously been shown to significantly improve its electronic mobility[11,21], presumably through the desorption of residue from the graphene surface. The ability to directly measure the mass of graphene resonators provides a tool to monitor this process, and to correlate electrical performance with the amount of residue. Figure 4a shows the conductance of graphene resonator as a function of gate voltage, measured after each of four successive ohmic heating steps. At each step, the conductance minimum (the Charge Neutrality Point, CNP) moves closer to $V_g = 0$, and the n-branch conductance increases while the p-branch decreases. Figures 4b-e show the resonant frequency as a function of gate voltage for the same device, in the as-fabricated state (Fig. 4b), and after each of the first three heating steps (Fig. 4c,d,e). The resonant frequency shifts upward at each step, consistent with a loss of mass; fitting the data to the membrane model shows that the two-dimensional density decreases monotonically, from 4.7 to 2.1 times that of graphene. The built-in strain in the sheet also changes slightly with each cleaning step. In this experiment, the device was damaged after the fourth step, preventing full removal of the PMMA residue. The room-temperature quality factor did not change appreciably during the entire process.

We next tested the response of a resonator to the addition of a known amount of mass. This was accomplished by evaporating pentacene[24] onto the device in a vacuum chamber, while measuring the device response *in situ*. The added mass was calibrated by a quartz crystal microbalance (QCM). Figure 5a shows the resonant frequency as a function of gate voltage for the as-fabricated device, which is again fit to yield the density and strain. The CNP is located at $V_g < -10V$. Figure 5b shows the corresponding data after deposition of 1.5 nm of pentacene. At high $|V_g|$, where there is a large gate-induced tension, the added mass results in the expected decrease in the resonant frequency. At low $|V_g|$, however, the resonant frequency upshifts, indicating that the added pentacene increases the built-in strain. Similar behavior is observed after a second 1.5nm evaporation step (not shown). The graphene was then ohmically heated to remove the adsorbed material. After cleaning (Fig. 5c), the resonant frequency shifts downward at low $|V_g|$ and shifts dramatically upward at high $|V_g|$, consistent with a decrease in both the mass and the built-in tension. An additional mode, whose nature is uncertain, is also seen, and the CNP is observable in the measured $V_g$ range. Finally, 1.5 nm of pentacene is again evaporated on the clean device (Fig. 5d). Once again, the frequency shifts downward at high $|V_g|$ and upward at low $|V_g|$, and the CNP moves to the left. This result suggests the possibility of using graphene resonators as multifunctional sensors by simultaneously tracking the resonant frequency and the CNP displacement[2]. The use of ohmic heating to regenerate the devices is also shown to be possible, although the process seems to damage the sheet in some cases. Improvements to the cleaning process may be needed in order to minimize the damage and extend the device lifetime. As expected, the gate-independent metal resonances remain largely unchanged during the entire process.

From the continuum model, the sheet mass density is found increase after each evaporation step by an amount equivalent to 1.5 nm of pentacene, in agreement with the thickness measured by the QCM. This provides strong confirmation of the model. In addition, in this device, ohmic heating is sufficient to remove essentially all of the residue, yielding a density close to that of pristine graphene (Fig. 5c). In this device, the built-in strain increases with the thickness of the adsorbed layer (Fig. 3b), indicating that adorbates can impart a tension to the suspended graphene sheet.

In analyzing the mass response of NEMS devices, one typically assumes that mass loading does not change the mechanics of the device. The frequency shift[16] is then simply $\frac{\Delta f}{f} \approx -\frac{1}{2}\frac{\Delta m}{m}$. An important result of this study is that the response of atomically thin resonators is not so simple, since the interaction between adsorbates and the sheet also must be considered. For instance, adsorbed mass can even shift the resonance *upward* in frequency when the tension is small, as is observed in our devices near $V_g = 0$. Using devices with large built-in tension should minimize the effects of this interaction. Alternatively, it may be possible to use this effect to discriminate between different types of adsorbates by the tension they impart to the sheet.

**Low temperature behavior**

To explore further the performance of the graphene resonators, we studied their behavior as a function of temperature. Upon cooling, the resonant frequency shifts upwards, and its tunability with $V_g$ decreases (Fig. 6a). This behavior is due to the geometry of our devices: when the temperature decreases, the suspended metal electrodes contract isotropically (see Supplementary Information), imparting tension to the graphene sheet. The frequency shifts are consistently observed in the four devices measured to date at low temperature, and are fully reversible upon warming, indicating that no slipping occurs between the graphene and the metal electrodes. The large temperature response shows the potential of using thermal effects for frequency tuning[25], or for direct drive of the resonance[26]. The built-in strain measured using the continuum model increases upon cooling, from $4.7 \times 10^{-4}$ at 295 K to $6.9 \times 10^{-4}$ at 250 K. The mass density of the resonator, on the other hand, remains essentially constant ($6.4 \rho_{graphene}$), as expected. At lower temperatures, the lack of measurable tunability prevents independent determination of both density and strain. However, the strain can be determined by using the density obtained at higher temperatures. At 125K, for instance, the strain is found to be $2.3 \times 10^{-3}$.

For practical use in MEMS/NEMS devices and other applications, accurate measurement of thermal expansion is important in order to enable modeling of thermal effects. In addition, this effect is of fundamental interest in graphene: as a two dimensional material, it possesses an unusual negative thermal expansion coefficient due to out-of-plane vibrational modes[27]. We used a finite element model to determine the shrinkage of the substrate and electrodes, in order to infer the thermal expansion coefficient of the graphene from the measured strain. When bulk values for the thermal expansion coefficients of gold and silicon are used, the model predicts that cooling from 295K to 250K should impart an additional strain $\Delta\varepsilon = 5.5\times10^{-4}$ to the graphene, which is larger than the observed increase. The difference is due to the negative thermal expansion coefficient of graphene, for which we derive a value $\alpha_{graphene} = -7.4\times10^{-6}/K$. This is larger in magnitude than predicted theoretically[28], but is in excellent agreement with recently reported experimental results[29]. The observed upshift of the resonant frequency of the gold electrodes is consistent with a value of the gold thermal expansion coefficient within 5% of the bulk value, and a derived $\alpha_{graphene}$ within 6% of that calculated above. These results clearly demonstrate the utility of resonant techniques to measure the thermal expansion of graphene and similar materials, and a more detailed study of this phenomenon is currently underway. In addition, we note that, with suitable choice of sample and electrode geometry, the thermally induced frequency shifts in graphene resonators can be engineered to be positive, negative, or zero.

In addition to the frequency shifts, we observe an increase in the quality factor with decreasing temperature. Figure 6b shows the temperature dependence of the dissipation $Q^{-1}$ for three devices, as measured by fitting the resonance peaks to a Lorentzian function in the linear response regime. These devices have varying electrical properties and levels of residual PMMA, but all three show remarkably similar behavior. Upon cooling from room temperature, $Q^{-1}$ drops quickly (roughly as $T^3$) and the quality factor reaches ~3000 at 100K. Below this temperature, the dissipation decreases more slowly, roughly as $T^{0.3-0.4}$. The quality factor reaches ~$10^4$ at 5 K (inset of Fig. 6b). In this regime, both the temperature dependence and magnitude are remarkably close to those recently reported for ultraclean carbon nanotubes[30] at lower temperatures (Q ~ $10^5$ at 0.01K, and $Q^{-1}$ ~ $T^{0.36}$ and from 0.01-1K). NEMS resonators made from nanocrystalline diamond[31] ($T^{0.2}$), GaAs[32] ($T^{0.25}$) and other bulk materials[33] also show similar temperature dependence at low T. In this study, samples of different size, electrical properties, and cleanliness show nearly identical behavior. The origin for this temperature dependence is still unknown.

**Conclusions**

Using the measured response to mass loading and the observed dynamic range, we estimate the mass sensitivity of the best sample (Q = 14,000 at low T) to be ~2 zg, with a detection bandwidth set by the lock-in amplifier integration time (300 ms). This is a somewhat larger mass than recently demonstrated using nanotubes[34,35], but could be reduced by improving the noise performance of the electrical detection system. Compared to nanotubes, graphene resonators have the advantage of more reproducible electrical properties and a larger surface area for capture of the incoming mass flux. Since the electrical properties of graphene are sensitive to adsorbed species[2], it should also be possible to achieve multifunctional devices that combine charge and mass sensing down to the single molecule level.

Compared to other materials commonly used for NEMS, graphene possesses an important and as-yet unexploited advantage: it can withstand ultrahigh strains, up to ~25% in nanoindentation experiments[5] and ~3-5% for micron-sized samples subjected to uniaxial strain[36,37]. In top-down fabricated NEMS, high resonant frequencies are achieved by reducing the device dimensions. This size reduction reduces both the output signal magnitude and the amplitude at the onset of nonlinearity, decreasing the dynamic range and making GHz-range transduction difficult. Micron-scale graphene devices, subjected to strains of order 1%, should achieve GHz operation while maintaining the robust signal levels observed in this study. An additional advantage is that the amplitude at the onset of nonlinearity increases with strain (Eq.

2), so that it should be possible to simultaneously increase the frequency and dynamic range (and therefore the mass sensitivity). It may be possible to achieve such large strains by integrating graphene with MEMS/NEMS structures.

**Methods**

Device fabrication

Graphene flakes (Toshiba Ceramics) are deposited on a $SiO_2$/Si substrate by mechanical exfoliation, optically located by their contrast[38] and subsequently confirmed to be monolayers with Raman spectroscopy[39]. Metal electrodes (Cr/Au) are patterned on the graphene by electron beam lithography. An optional second step of lithography and oxygen plasma etching can be used to shape the graphene after patterning of electrodes. The devices are then suspended by etching the $SiO_2$ epilayer using buffered oxide etchant (BOE) followed by critical point drying to suspend the graphene between the electrodes[11]. We have found that BOE etches uniformly underneath the graphene layer, rather than isotropically from the edge as for most materials[40]. This permits suspension of devices with a uniform gap below the graphene, even for sheets with dimensions over 1 micron. The resulting wide suspended graphene sheets are low-impedance devices, which is helpful in matching to high-frequency electronics. Lithographic control also allows narrow ribbon resonators[41], as shown in Fig. 1b. The combination of optical thickness identification and lithographic patterning gives us full three-dimensional knowledge of the device geometry. Although the oxide layer below the graphene in the area covered by the electrode is removed in the etching process, the adhesion between graphene and metal still provides good mechanical clamping.

Electromechanical Mixing

The method implemented here is derived from previous studies of carbon nanotube resonators[14], with minor modifications detailed in the Supplementary Information. The suspended graphene is actuated by the potential difference between the sheet and the silicon substrate (gate). A DC voltage $V_g$ is applied to the gate electrode to adjust the tension in the sheet, and an RF voltage $\delta V_g^f$ sets it into motion at frequency $f$. A second RF signal $\delta V_{sd}^{f \pm \Delta f}$ is applied to the drain electrode, so that a mixed-down current appears across the device at the intermediate frequency $\Delta f$ with amplitude:

$$I^{\Delta f} = \frac{dG}{dV_g} \delta V_{sd}^{f \pm \Delta f} (\delta V_g^f + V_g \frac{\delta z^f}{z}) \qquad (3)$$

where $G = \partial I / \partial V_{sd}$ is the conductance of the sheet, $\delta z^f$ is the amplitude of vibration at the driving frequency and $z$ is the graphene-substrate distance. The magnitude of this current depends on the modulation of the conductance with gate voltage $dG/dV_g$, and is detectable for drain and gate signals in the millivolt range. The conductance swing per micron width of graphene is typically > 10 µS/V at room temperature, a consequence of the large contact area, low contact resistance and high mobility of monolayer graphene, and can be improved by increased gate efficiency[4,11]. The first term in parentheses is purely electronic and results in a smooth background, while the second term results from the mechanical vibration of the sheet and peaks at the resonant frequency. All the measurements are carried out in a vacuum chamber (<10$^{-5}$ Torr).

Acknowledgement

We thank Mingyuan Huang for useful discussions, Hugen Yan for Raman spectroscopy, and Vincent Lee for evaporator setup. This work is supported by DARPA Center on Nanoscale Science and Technology for Integrated Micro/Nano-Electromechanical Transducers (iMINT,

Grant No HR0011-06-1-0048, Dr. D. L. Polla, Program Manager), the National Science Foundation (Grant CHE-0117752), the W. M. Keck Foundation, and Microsoft Project Q.

Author contributions

C.C., S.R. and J.H. designed the experiment. C.C. and W.K. performed the fabrication, S.R. and C.C. performed the experiments and analyzed data, K.B. assisted with the fabrication, measurement and data analysis. I.K. provided assistance with mass sensing. C.C., S.R. and J.H. co-wrote the paper. P.K., H.L.S., and T.F.H. provided materials and equipment. All authors discussed the results and commented on the manuscript.

Additional information

Supplementary information accompanies this paper at www.nature.com/naturenanotechnology. Reprints and permission information is available online at http://npg.nature.com/reprintsandpermissions/. Correspondence and requests for materials should be addressed to J.H.

3   Mohanty, N. & Berry, V., Graphene-Based Single-Bacterium Resolution Biodevice and DNA Transistor: Interfacing Graphene Derivatives with Nanoscale and Microscale Biocomponents. *Nano Lett.* **8**, 4469-4476 (2008).

4   Meric, I. et al., Current saturation in zero-bandgap, top-gated graphene field-effect transistors. *Nature Nanotech.* **3**, 654-659 (2008).

5   Lee, C., Wei, X. D., Kysar, J. W., & Hone, J., Measurement of the elastic properties and intrinsic strength of monolayer graphene. *Science* **321**, 385-388 (2008).

6   Mamin, H. J. & Rugar, D., Sub-attonewton force detection at millikelvin temperatures. *Appl. Phys. Lett.* **79**, 3358-3360 (2001).

7   LaHaye, M. D., Buu, O., Camarota, B., & Schwab, K. C., Approaching the quantum limit of a nanomechanical resonator. *Science* **304**, 74-77 (2004).

8   Yang, Y. T. et al., Zeptogram-scale nanomechanical mass sensing. *Nano Lett.* **6**, 583-586 (2006).

9   Bunch, J. S. et al., Electromechanical resonators from graphene sheets. *Science* **315**, 490-493 (2007).

10  Garcia-Sanchez, D. et al., Imaging mechanical vibrations in suspended graphene sheets. *Nano Lett.* **8**, 1399-1403 (2008).

11  Bolotin, K. I. et al., Ultrahigh electron mobility in suspended graphene. *Solid State Commun.* **146**, 351-355 (2008).

12  Stolyarova, E. et al., Observation of Graphene Bubbles and Effective Mass Transport under Graphene Films. *Nano Lett.* **9**, 332-337 (2009).

13  Witkamp, B., Poot, M., & van der Zant, H. S. J., Bending-mode vibration of a suspended nanotube resonator. *Nano Lett.* **6**, 2904-2908 (2006).

14  Sazonova, V. et al., A tunable carbon nanotube electromechanical oscillator. *Nature* **431**, 284-287 (2004).

15  Knobel, R. G. & Cleland, A. N., Nanometre-scale displacement sensing using a single electron transistor. *Nature* **424**, 291-293 (2003).

16  Ekinci, K. L. & Roukes, M. L., Nanoelectromechanical systems. *Rev. Sci. Instrum.* **76**, 061101 (2005).

17  Lassagne, B., Garcia-Sanchez, D., Aguasca, A., & Bachtold, A., Ultrasensitive Mass Sensing with a Nanotube Electromechanical Resonator. *Nano Lett.* **8**, 3735-3738 (2008).

18  Atalaya, J., Isacsson, A., & Kinaret, J. M., Continuum Elastic Modeling of Graphene Resonators. *Nano Lett.* **8**, 4196-4200 (2008).

19  Bunch, J. S. et al., Impermeable atomic membranes from graphene sheets. *Nano Letters* **8**, 2458-2462 (2008).

20  Yakobson, B. I., Brabec, C. J., & Bernholc, J., Nanomechanics of carbon tubes: Instabilities beyond linear response. *Phys. Rev. Lett.* **76**, 2511-2514 (1996).

21  Moser, J., Barreiro, A., & Bachtold, A., Current-induced cleaning of graphene. *Appl. Phys. Lett.* **91**, 163513 (2007).

22  Ishigami, M. et al., Atomic structure of graphene on $SiO_2$. *Nano Lett.* **7**, 1643-1648 (2007).

23  Postma, H. W. C., Kozinsky, I., Husain, A., & Roukes, M. L., Dynamic range of nanotube- and nanowire-based electromechanical systems. *Appl. Phys. Lett.* **86**, 223105 (2005).

24  Dimitrakopoulos, C. D. & Malenfant, P. R. L., Organic thin film transistors for large area electronics. *Adv. Mater.* **14**, 99-117 (2002).

25  Jun, S. C. et al., Electrothermal tuning of Al-SiC nanomechanical resonators. *Nanotechnology* **17**, 1506-1511 (2006).

26  Bargatin, I., Kozinsky, I., & Roukes, M. L., Efficient electrothermal actuation of multiple modes of high-frequency nanoelectromechanical resonators. *Appl. Phys. Lett.* **90**, 093116 (2007).
17

Figure 1 **Device and experimental setup**. **a**, Scanning electron microscope (SEM) image of several resonators made from single monolayer graphene flake. **b**, SEM image of suspended graphene nanoribbon, lithographically patterned to a width of 200 nm before suspension. **c**, Schematic of suspended graphene. The SiO$_2$ under the entire graphene flake is etched evenly, including the contact region. **d**, Diagram of electronic circuit. A carrier RF signal $\delta V_{sd}^{f \pm \Delta f}$ is applied to the drain, and an RF drive signal $\delta V_g^f$ together with DC bias $V_g$ is applied to the gate. The current through the graphene is detected by a lock-in amplifier after pre-amplification, at an intermediate frequency $\Delta f$ of 1 kHz.

Figure 2 **NEMS properties of graphene resonators. a**, Mixed-down current $\delta I^{\Delta f}$ versus frequency for device 1 (3μm wide, 1.1 μm long), taken at $V_g = -7V$ from **b** (dashed black vertical line), with RF drive amplitude $\delta V_g = 19 mV$ and RF bias $\delta V_{sd} = 110 mV$. The graphene resonance (I) appears at 65 MHz. Resonances of metal beams (II) are also visible below 25 MHz. Inset: the graphene resonance at low driving power, and Lorentzian fit (red line) with Q=125. **b,c** $\delta I^{\Delta f}$ in color scale as a function of both driving frequency and DC gate voltage $V_g$ for device 1 and device 2 (200 nm wide, 1.8 μm long). The graphene resonances, visible as parabola-shaped features, are highly tunable with gate voltage for both devices. In **b**, besides graphene resonances (I), resonances from the metal clamps are observed, both as non-tunable modes (II) and as avoided crossing points (arrows). These modes are absent in device 2, which is much narrower. **d**, $\delta I^{\Delta f}$ versus driving frequency for different drive amplitudes ($\delta V_g$), with $V_g = -7V$, $\delta V_{sd} = 7.5 mV$. The estimated vibration amplitudes ($\delta z$) are given for each $\delta V_g$ (see Supplementary Information), the onset of nonlinearity effects corresponds to $\delta V_g = 44 mV$, with estimated vibration amplitude of 1.1 nm. Beyond this linear regime, the amplitude estimation is not valid.

Figure 3 **Modeling of device behavior. a**, Measured resonant frequency versus $V_g$ for two devices, fit using a membrane model, with the values of the fitting parameters shown. **b**, Built-in strain vs. normalized mass density. The solid squares show data for 11 as-fabricated devices. The open triangles show data for the device cleaned by ohmic heating described in figure 4. The open squares show data for the pentacene deposition (red arrows) and the removal of pentacene (dashed line) by annealing process described in figure 5. The dashed black line shows the mass density of pristine graphene.

Figure 4 **Removal of mass by ohmic heating**. **a**, Room-temperature conductance ($V_{sd}$=1 mV) versus gate voltage for a device before (black curve) and after ohmic-heating cycles to $V_{sd}$=1.5 V (red), 1.6 V (green), 1.8 V (blue), and 3 V (pink). Successive curves show decreased conductance at negative gate voltage (black arrow), increased conductance at positive gate voltage, and conductance minima closer to zero gate voltage. **b-e**, Resonant response (frequency-gate voltage response plotted as a derivative in color scale), for the as-fabricated device, and after the 1.5 V, 1.6 V, and 1.8 V, respectively. The derived strain and density for each step are shown, and confirm that residue is removed. Due to damage to the device, it was not possible to obtain a resonance curve after the 3 V annealing step.

Figure 5 **Effects of Mass Loading**. **a**, Frequency-gate voltage response plotted as a derivative in color scale, of as-fabricated device. **b**, Response of the same device after evaporation of 1.5nm pentacene. **c**, Response after cleaning by ohmic heating. **d**, Response after deposition of another 1.5 nm pentacene. The derived strain and density for each step are shown. Bottom panels show the DC conductance as function of $V_g$ for each step.

Figure 6 **Temperature dependence**. **a**, Frequency-gate voltage responses at different temperatures, showing upshift of resonances and decreasing tunability. The origin of the peak splitting at 250 K is not clear. **b**, Temperature dependence of energy dissipation ($Q^{-1}$). Different

symbols indicate data from different devices. One of them (red circles) shows increasing resistance (up to ~1MΩ) upon cooling while the others retain resistance in kΩ range. The dashed line shows $T^{0.36}$ behavior reported for similar measurements on carbon nanotubes[30]. Inset: resonant peak, showing quality factor of 14,000 at 5 K.

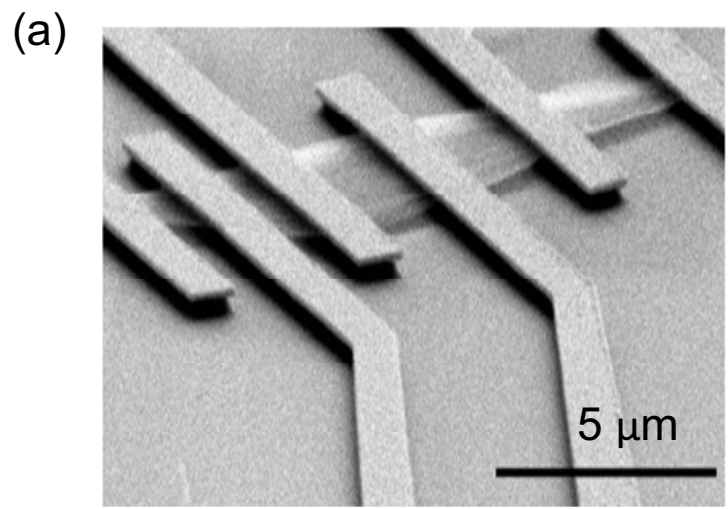
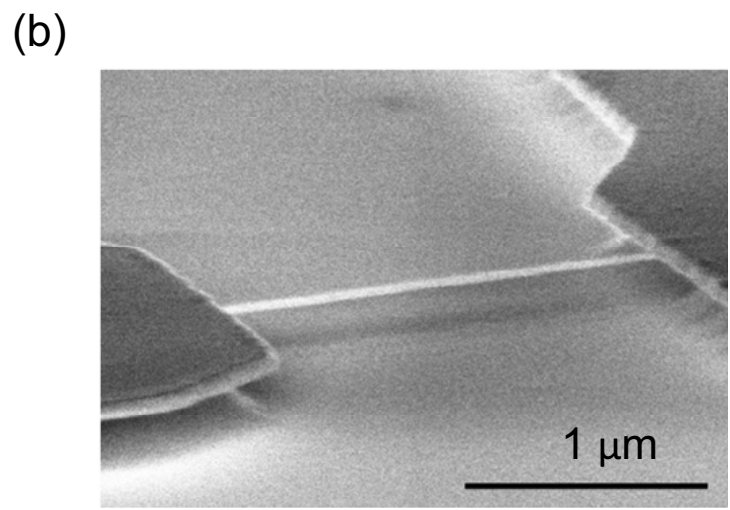
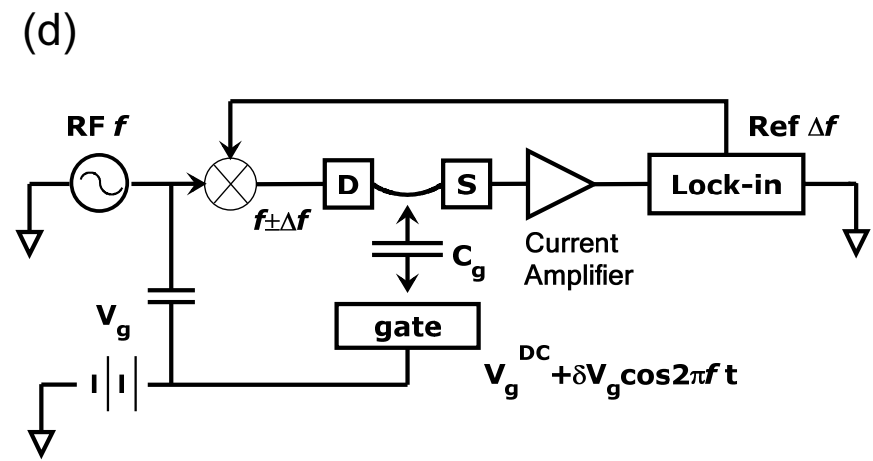
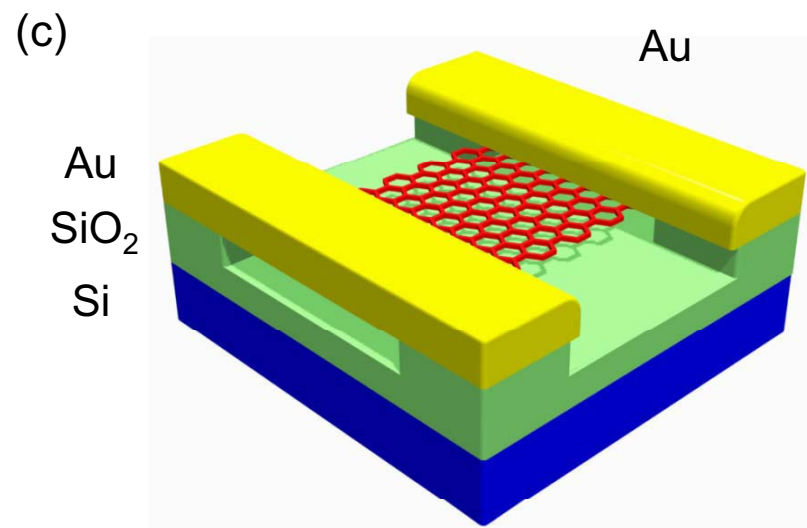

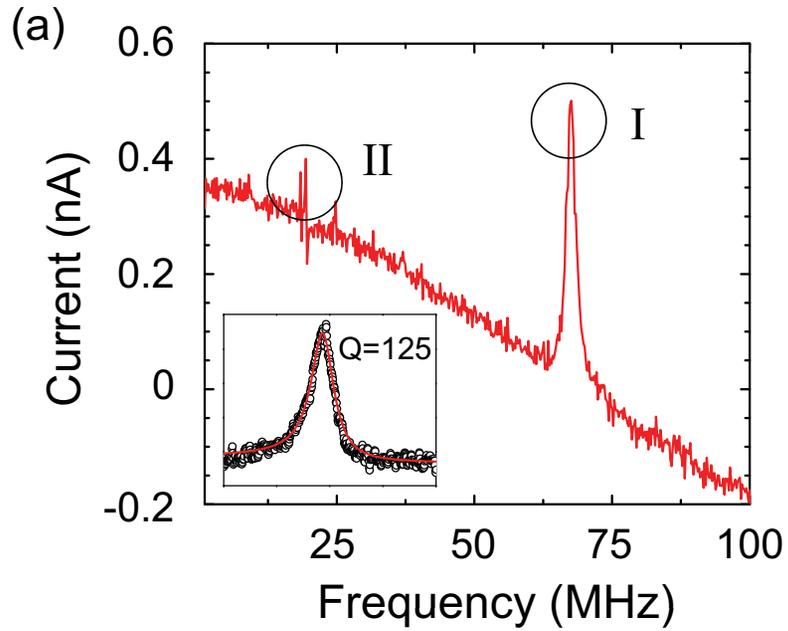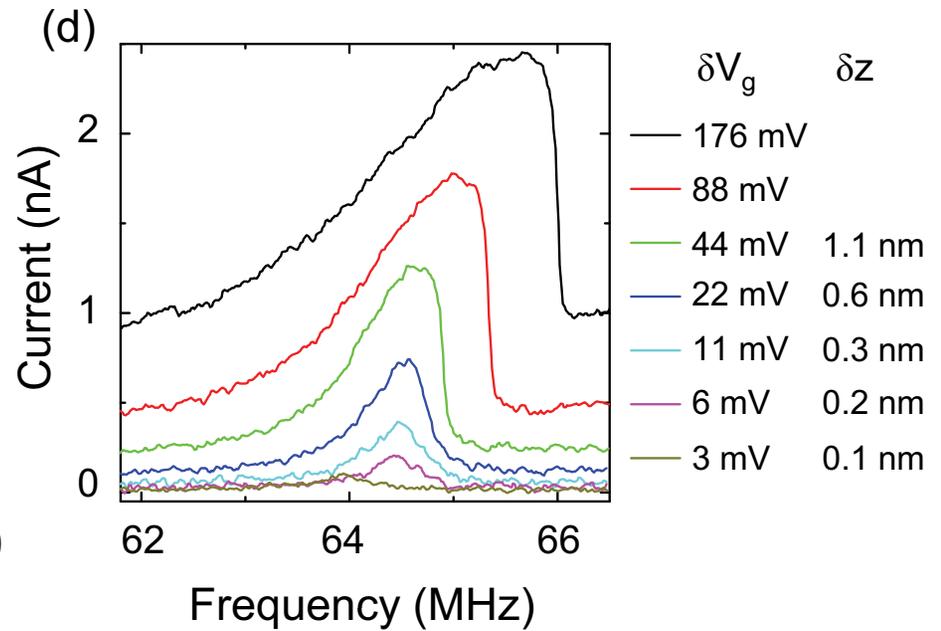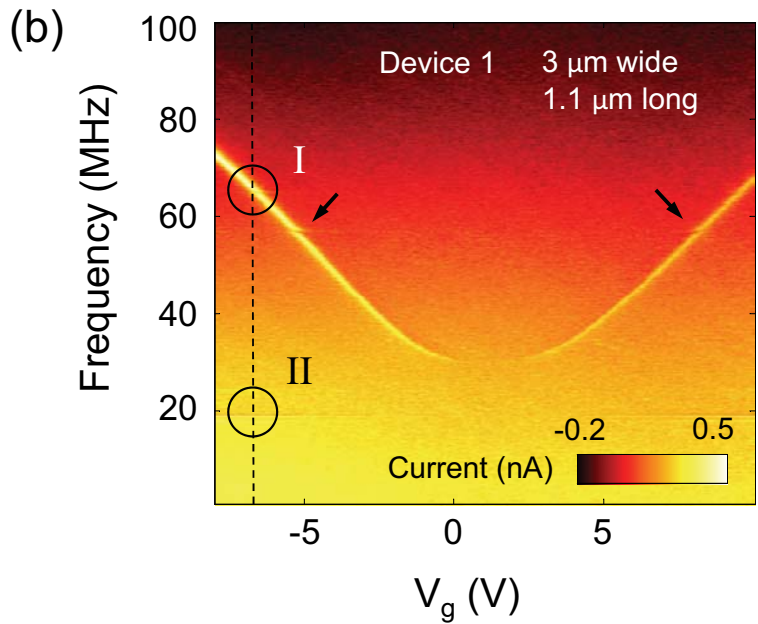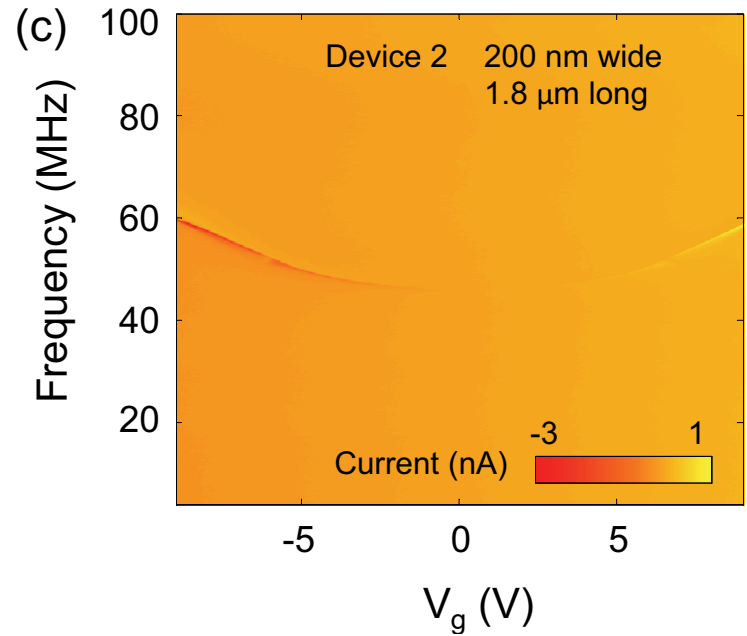

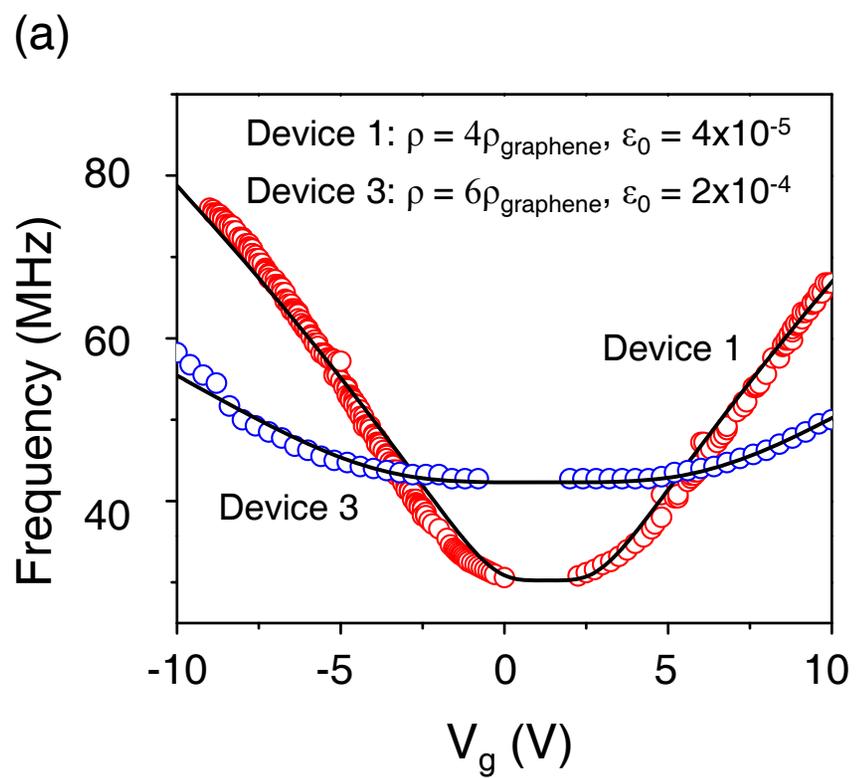 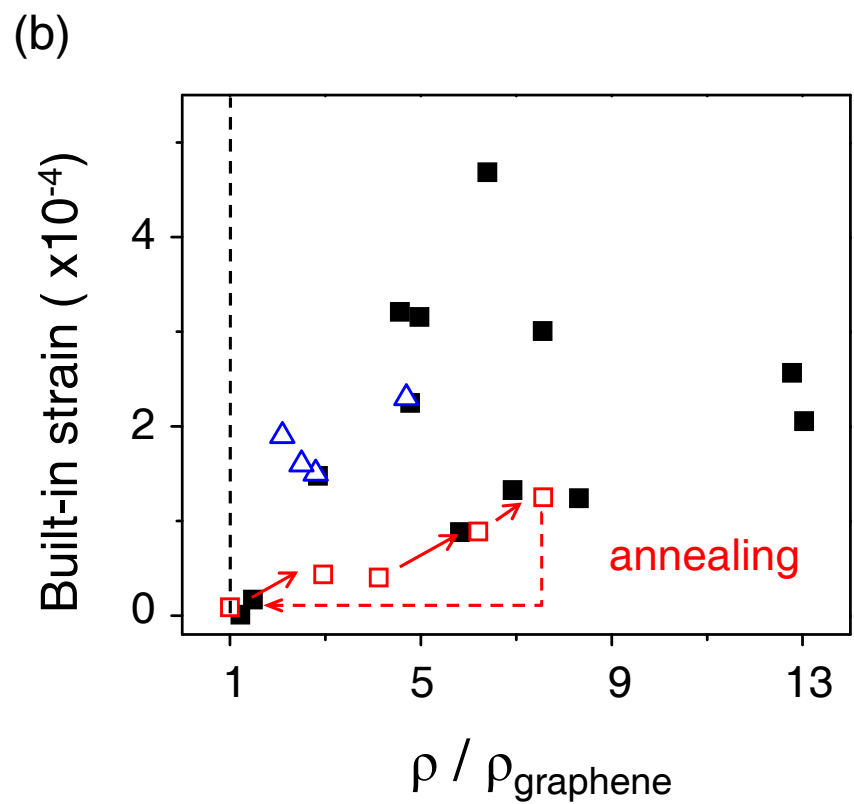

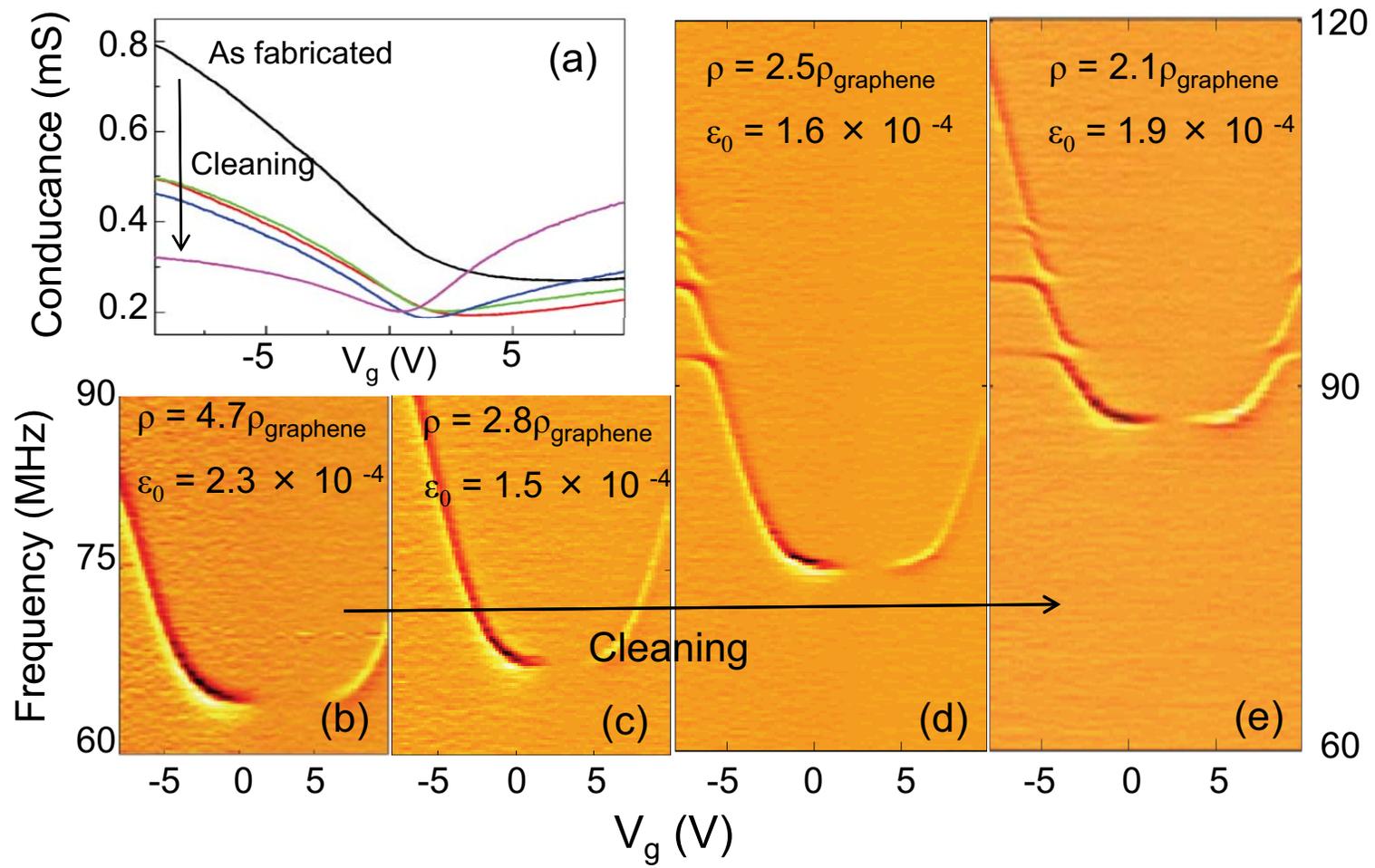

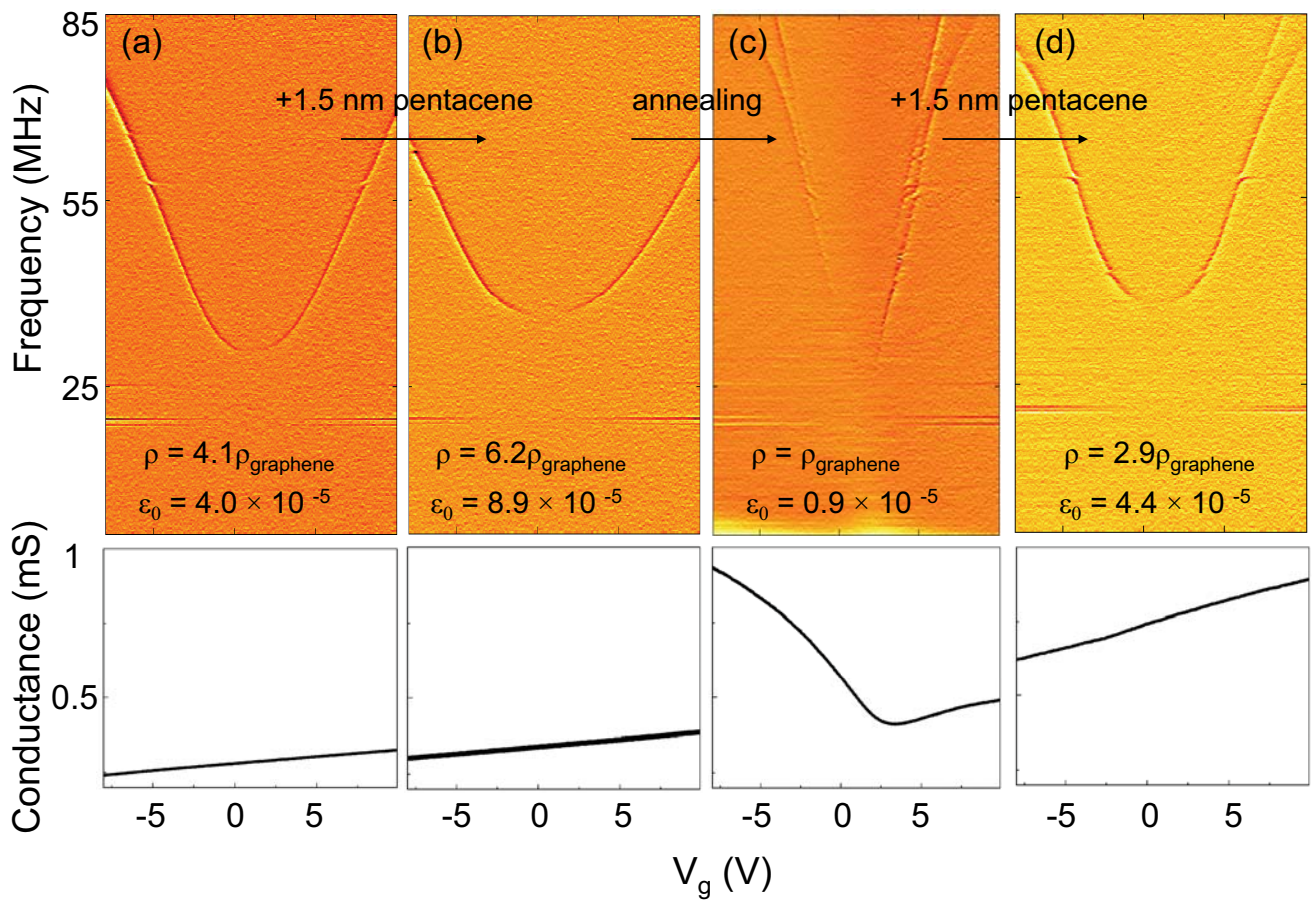

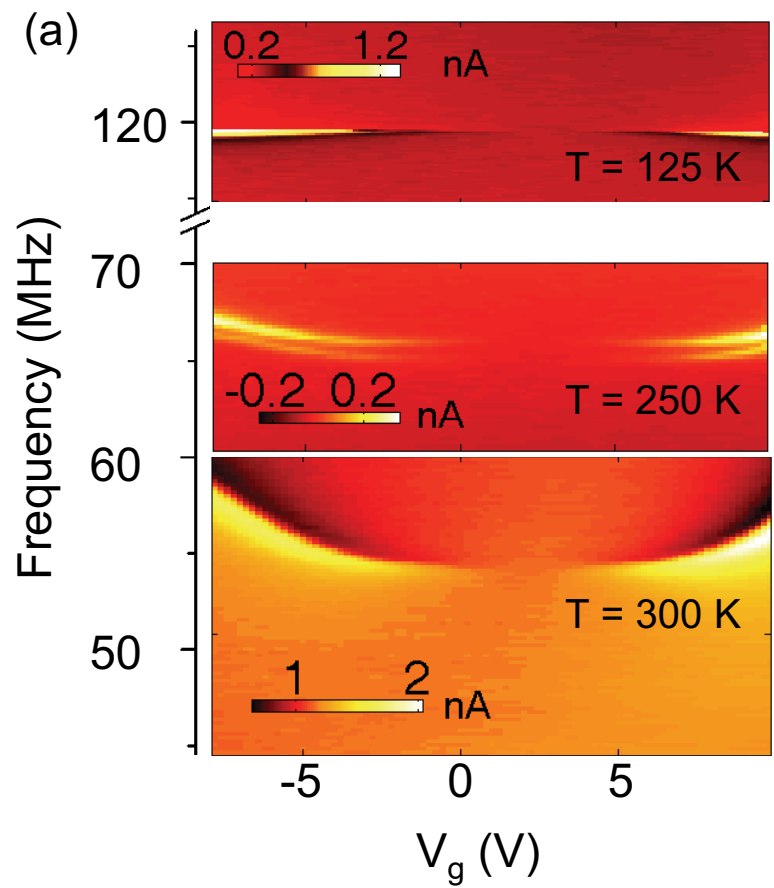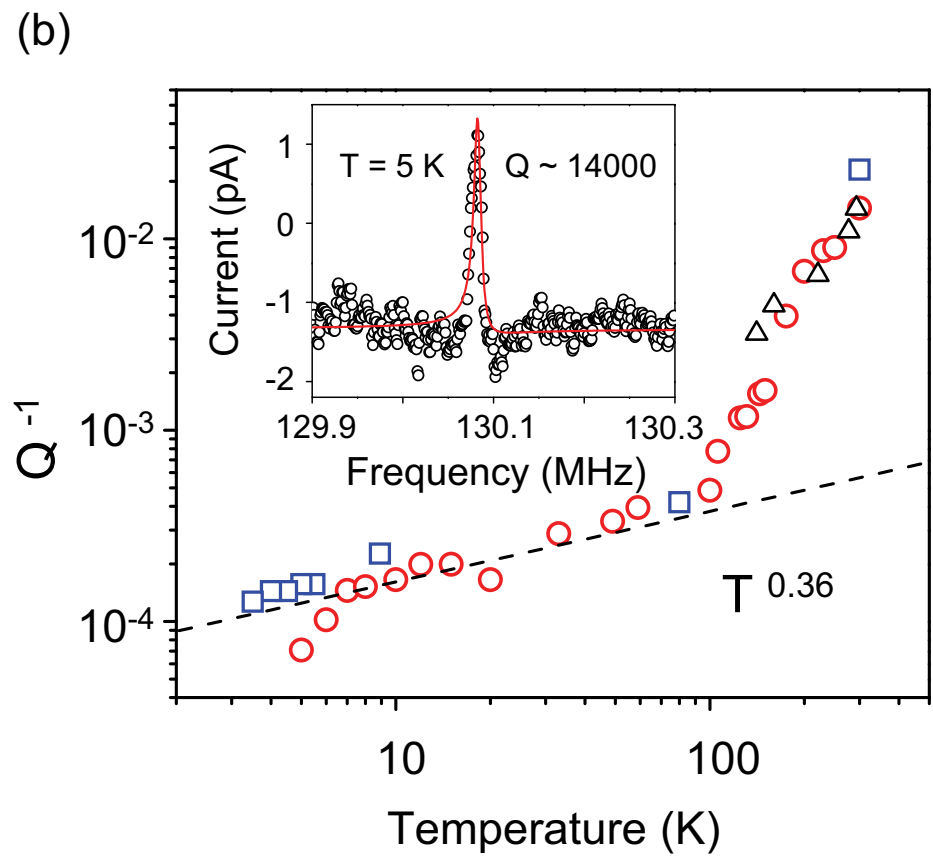

# Supplementary Information for

## *Performance of Monolayer Graphene Nanomechanical Resonators with Electrical Readout*

Changyao Chen, Sami Rosenblatt, Kirill I. Bolotin, William Kalb, Ioannis Kymissis, Horst L. Stormer, Philip Kim, Tony F. Heinz, James Hone*

*e-mail: jh2228@columbia.edu

### A. Sample identification

Graphene flakes are deposited by mechanical exfoliation and patterned by electron beam lithography. After electrode deposition and $SiO_2$ removal, samples are transferred and dried with a critical point dryer (model 13200J-AB from SPI-DRY) to avoid surface tension during the phase transition that causes suspended graphene to collapse. Samples are checked by atomic force microscopy (AFM) to verify suspension. Raman spectroscopy is then performed on the suspended region; the width of the 2D mode identifies the number of layers (Fig. S1). This mode shows no difference before and after graphene suspension.

### B. Electrical measurement system

Samples are wire bonded and mounted onto a chip carrier. Coaxial cables with a characteristic impedance of 50 Ω are soldered directly to the chip carrier. Each input (drain and gate) is terminated with a matching load consisting of a 50Ω resistor in series with a 100 nF capacitor to ground (Fig. S2a). This load allows for simultaneous DC and RF biasing (realized by means of a wideband bias tee, ZFBT-4R2GW+ from Minicircuits) while minimizing RF reflections. Each output (source) is terminated with a 100 nF capacitor to ground, which not only is used as a low-pass filter with corner frequency below 30 kHz, but also serves the purpose of anchoring the RF ground to eliminate ringing in the output; in the absence of the capacitor, the mixing current undergoes large fluctuations as a function of frequency, as the output cable becomes a quarter-wave transformer with variable impedance. The capacitance between the contact pads ($120\mu m \times 120\mu m$, with 300 nm $SiO_2$ dielectric) and the underlying silicon gives a cutoff frequency of ~1 GHz. The frequency mixer used to generate the sidebands can be applied either to the drain or gate side with equal results. No sideband rejection is employed, effectively doubling the amount of current generated by one sideband at the output. Power control is obtained by a combination of fixed attenuators and a digital step attenuator (model ZX-76 from Minicircuits).

Due to the phase difference between the mechanical oscillation and the electrical signal, we usually see asymmetric mixed-down current lineshapes instead of the expected Lorentzian peak[1]. We correct the phase of the electrical signal by simply changing the length of the coaxial transmission line (BNC cable) to either drain or gate, until a Lorentzian resonance lineshape is observed (Fig. S2b). In the 10-100 MHz regime, this correction technique is possible because a quarter wavelength is of the order of a few feet (at larger frequencies, commercial delay lines may be used).

**C. Background mixing current and Charge Neutrality Point (CNP)**

The mixing current has two contributions: a background current $I_{BG}$ which exists even in non-suspended samples and a resonant term $I_{peak}$ which is proportional to the vibration amplitude. The background mixed-down current is given by:

$$I_{BG} = \frac{dG}{dV_g} \delta V_{sd} \delta V_g \qquad (S1)$$

where $G$ is the conductance, $V_g$ is the DC gate voltage, and $\delta V_{sd}$ and $\delta V_g$ are the RF voltages applied to the drain and gate, respectively. Excellent agreement is observed between the mixed-down current and the expected value from the above equation at low frequencies (compared to the resonance but still significantly larger than the modulation frequency $\Delta f$), as seen in Figs. S2c and S2d. The measured mixing current at the CNP is zero as expected. However, we often find it located beyond +/- 10 V for as-fabricated devices, resulting in a nearly constant slope ($dG/dV_g$). Joule heating typically brings the CNP to the vicinity of $V_g = 0$, and additional mass deposition drives it away, due to charge doping induced by the evaporated material.

**D. Vibration amplitude estimation and nonlinear drive**

In the harmonic regime, the vibration amplitude has a Lorentzian lineshape. Assuming phase correction has been applied, the mixed-down current will follow the vibration, and the maximum amplitude of vibration will be given by:

$$\delta z_{max} = \frac{1}{2} Z_0 I_{peak} / (V_g \delta V_{sd} \frac{dG}{dV_g}) \qquad (S2)$$

where $I_{peak}$ is the current on resonance (excluding background) obtained from the Lorentzian fit, $\delta V_{sd}$ is the RF drain voltage, $V_g$ is the DC gate voltage, and $Z_0$ is the distance between the graphene plane and the substrate[1]. The factor of 2 comes from the use of two sidebands ($f + \Delta f$ and $f - \Delta f$) generated by the mixer in our setup.

If the sheet is driven beyond the harmonic regime, the vibration amplitude initially tops off and the current peak broadens. It then becomes hard to fit the peak with a simple Lorentzian lineshape and estimate the vibration amplitudes with precision, but it is clear that the quality factor worsens and that the oscillation stops increasing. Experimentally, we accomplish nonlinear studies by applying low power to the mixer input while applying large power to the other input. Given that commercial mixers may allow as much of ~ 20 dB at the main frequency, care must be taken when trying to measure the response to a large signal: if the amount of direct power appearing at the mixer exceeds that applied by the other input at the main frequency, then the mixer will drive both the oscillation and the electrical response. Because of the potential difference between drain and gate, large $\delta V_{sd}$ may also effectively drive the suspended graphene into resonance when $\delta V_g$ is small. With large $\delta V_{sd}$, we observe nonlinear lineshapes and hysteresis of the frequency sweep (Fig. S3).

**E. Assignment of gold and graphene resonances**

The 2D structure of graphene determines that its mechanical resonant frequency is highly sensitive to tension, which results in high tunability with external tension, as described by Eq. (1) in main text. In contrast, the resonant frequency of the suspended gold beam is not expected to be changed by relatively small external tension due to the large bending rigidity, and is given by (doubly clamped beam)[2]:

$$f_{gold} = 1.03 \frac{t}{L_{gold}^2} \sqrt{\frac{E}{\rho}} \quad (S3)$$

where $t$ is the thickness of the suspended metal beam (typical 80 nm Au on top of 1 nm Cr), $L_{gold}$ is the length of the suspended section of the gold electrode (slightly larger than width of graphene because of the isotropic etching of $SiO_2$), and $E$ and $\rho$ are the Young's modulus and density of bulk gold. Figure S4a shows the measured gold resonances as a function of $t/L_{gold}^2$. There is good agreement with Eq. (S3). We attribute the deviations from the theoretical prediction to the random built-in tension generated during the fabrication process. Figure S4b shows the measured resonant frequency of a series of 6 devices of different length fabricated from the same graphene flake, measured at high $V_g$ to minimize the effect of uncontrolled built-in tension. The frequency varies roughly as $1/L_{graphene}$, in agreement with Eq. (1). This scaling also shows that resonances in the GHz regime should be achievable for device lengths in the ~100 nm range.

**F. Continuum mechanics model**

Because the bending stiffness of the monolayer graphene is negligible[3], the graphene resonator can be treated as a doubly clamped membrane. Furthermore, we simplify our analysis to a 1D string with length $L$ (Fig. S5). The bonding between graphene and electrode provides good adhesion[4], and both AFM and SEM images after measurement confirm that the graphene sheets do not collapse. The electrostatic force between the sheet and the back gate is given by $F = \frac{1}{2}\frac{\partial C_g}{\partial z}V_g^2$, where $C_g$ is capacitance between the gate and the graphene (we assume a parallel-plate capacitor with spacing of 300 nm[ref kirill paper]) and $\frac{\partial C_g}{\partial z}$ is the spatial derivative. If we approximate the shape of the graphene under the electrostatic force as a circular arc with radius $r$, then the total tension in the string approximation can be treated as $T = T_0 + T_e = F/rL$, where the tension $T$ includes the initial tension $T_0$ and the external tension $T_e$ induced by the electrostatic force (Fig. S5). The external tension $T_e$ is a result of elastic elongation of the string[3], and therefore $T_e = wEt\Delta L/L$, where $w$ and $t$ are the width and thickness of the device, $E$ is the Young's modulus. When $\Delta L \ll L$, we have $T_e = wEtL^2/24r^2$, and finally $T_e(T_e+T_0)^2 = wEtF^2/24$, yielding

$$T = \frac{T_0}{3} + \sqrt[3]{\frac{a+\frac{2}{27}T_0^3 + \sqrt{a^2+\frac{4}{27}aT_0^3}}{2}} + \sqrt[3]{\frac{a+\frac{2}{27}T_0^3 - \sqrt{a^2+\frac{4}{27}aT_0^3}}{2}} \quad (S4)$$

where $a = \frac{1}{48}wEt(\frac{\partial C_g}{\partial z})^2 V_g^4$. Combined with equation (1) from the main text, we have an analytic expression to predict the resonant frequency of a graphene resonator under electrostatic force. Here, we corrected the offset of $V_g$ due to the trapped charge. When $T_e \gg T_0$ (at large $V_g$), the extra tension varies as $V_g^{4/3}$, so that the frequency varies as $V_g^{2/3}$. Fits to the measured data using this model with mass density and initial strain as free parameters provide excellent agreement with the experimental data.

When there is resist residue (Polymethyl methacrylate, PMMA) or pentacene on top of the graphene, we can treat the combination as a multilayer stack made from uniform films. Its equivalent two-dimensional elastic stiffness is[5]

$$\overline{Et} = \sum_{m=1}^{n} \overline{E_m} t_m \quad (S5)$$

where $\overline{E_m}$ and $t_m$ are the elastic modulus and thickness of $m^{th}$ layer, respectively. Since the Young's modulus of graphene[6] is three orders of magnitude larger than that of pentacene or PMMA (order of GPa), the equivalent elastic stiffness is dominated by the contribution coming from the graphene. The measured two-dimensional elastic stiffness of graphene (342 N/m) gives an effective Young's modulus of 1.02 TPa when a thickness of 0.335 nm is used.

## G. Thermal expansion coefficient of graphene resonator

When the operating temperature is lowered, we observe consistent increase of all resonant frequencies, as well as decrease of tunability over the same $V_g$ range (Fig. 6a). The upshift of resonant frequency at small $V_g$ implies an increase of built-in tension across the suspended graphene sheet, in turn minimizing the effect from external tension induced by $V_g$. Degradation of the tunability is therefore expected. This increase of tension comes from the isotropic contraction of suspended metal contacts (Fig. S6a), and is reversible upon thermal cycling. Although the contraction of the metal beam is not uniform due to the clamping to the oxide, we can consider it as equivalent to uniform shrinkage along the graphene resonator direction within this small deformation range. Since the suspended graphene is only anchored to metal contacts, which act like doubly clamped beams on the substrate, we are able to extrapolate the thermal expansion coefficient of the graphene resonator using the thermal expansion coefficient of gold (bulk)[7] and silicon[8].

**H. Mass sensitivity**

The minimum resolvable mass is determined as:

$$\delta m = \frac{\partial M}{\partial f} \delta f \times 10^{-DR/20} \approx -\frac{2M_{eff}}{Q} \times 10^{-DR/20} \tag{S6}$$

where $M = 7.4 \times 10^{-16} g$ is the mass of the graphene per $\mu m^2$, $Q \approx 14000$ is the quality factor, $DR = 40 dB$ is the dynamic range at T = 5K, thus this yields the result of 1 zg/μm².

The mass sensitivity $S_m^{1/2}$ is given by:

$$S_m^{1/2} = \frac{\partial M_{eff}}{\partial f_0} \frac{\partial f}{\partial I} S_n^{1/2} \tag{S6}$$

where $M_{eff} = 1.5 \times 10^{-15} g$ is the mass of the resonator, and $(\frac{\partial I}{\partial f}) \approx (\frac{\delta I}{\delta f_0}) = 1.24 \times 10^{-16} A/Hz$ is an approximation of current change when the device is swept through the resonance. From Fig 6b, $(\frac{\partial I}{\partial f})$ is the slope of the response function. $S_n^{1/2} = 4.12 \times 10^{-15} A/\sqrt{Hz}$ is the current noise spectral density of the measurement. Overall, the resulting mass sensitivity is $7.6 \times 10^{-22} g/\sqrt{Hz}$.

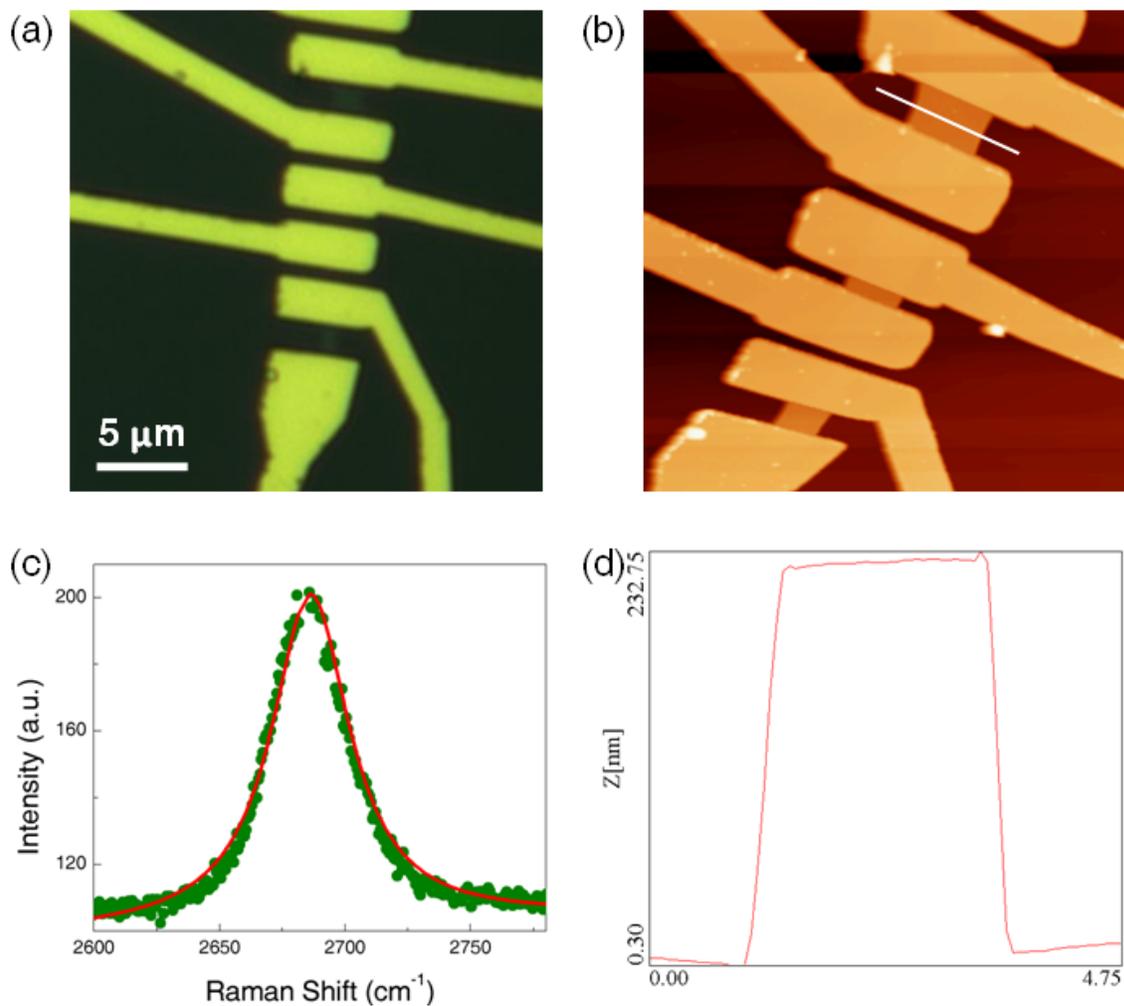

**Figure S1. a**, Optical microscope image of suspended graphene. **b**, AFM image of the same sample in **a**. **c**, Raman spectrum of the same sample in **a**, showing the lineshape of 2D peak (green dots) with Lorentzian fit (red line). **d**, Height profile corresponding to the white line section in **b**.

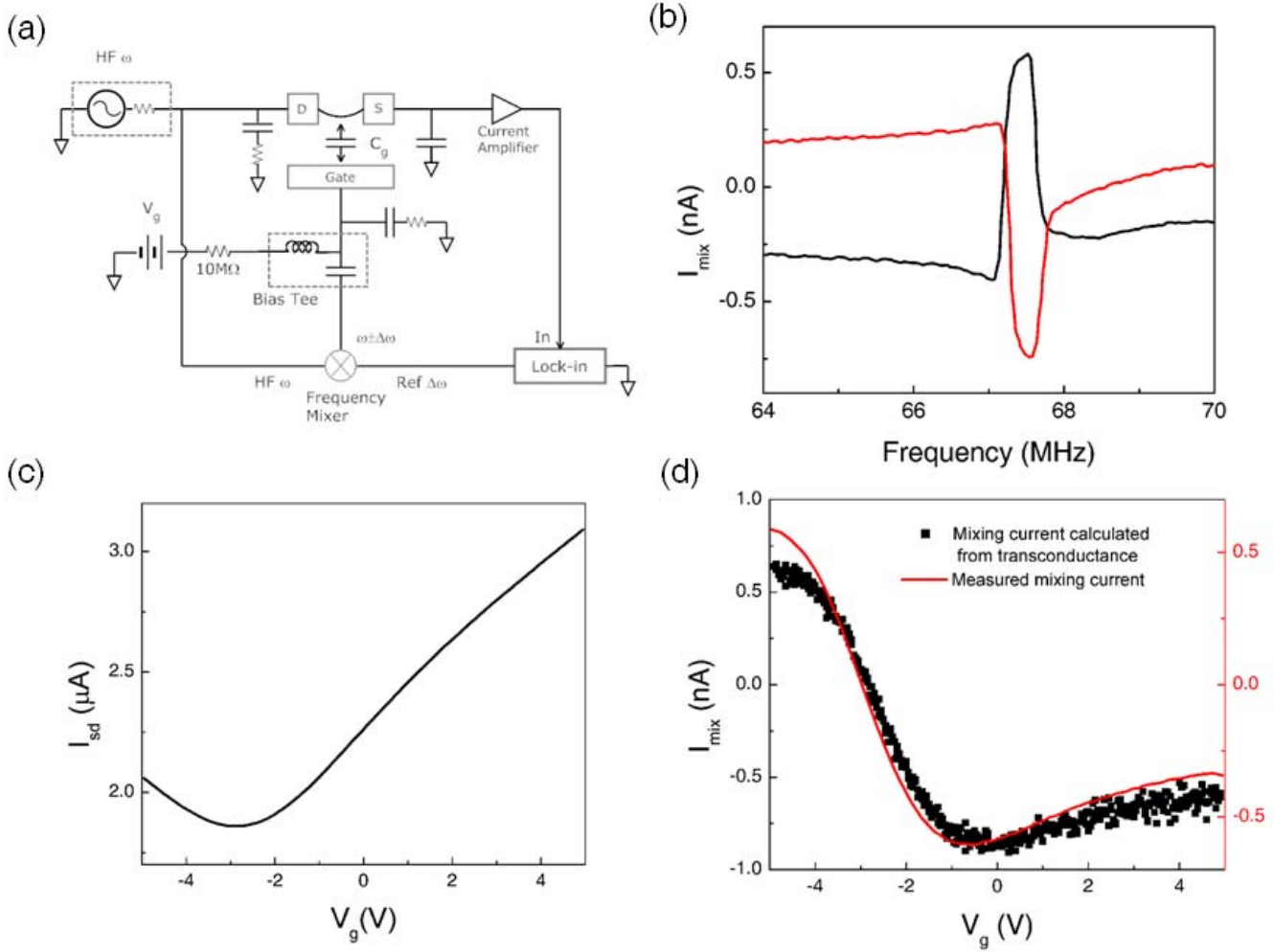

**Figure S2**. **a**, Electrical circuit setup with matching loads at each connection. **b**, Resonant peak with different lengths of transmission line (Pomona 2249-C, cable length added to drain side is 2.5 feet), showing different line shape due to phase change. Amplitudes (peak values): $\delta V_g = 14 mV$, $\delta V_{sd} = 100 mV$. The calculated phase change at 67 MHz is 170°. **c**, Gate sweep with $V_{sd} = 10 mV$, CNP is close to $V_g = -3V$. **d**, Mixing current measured from the same device shown in **c**, operated at 1 MHz, $\delta V_{sd} = 5 mV$, $\delta V_g = 100 mV$. Black points show predicted values of mixing current according to equation (S1). All mixing currents are recorded in rms values.

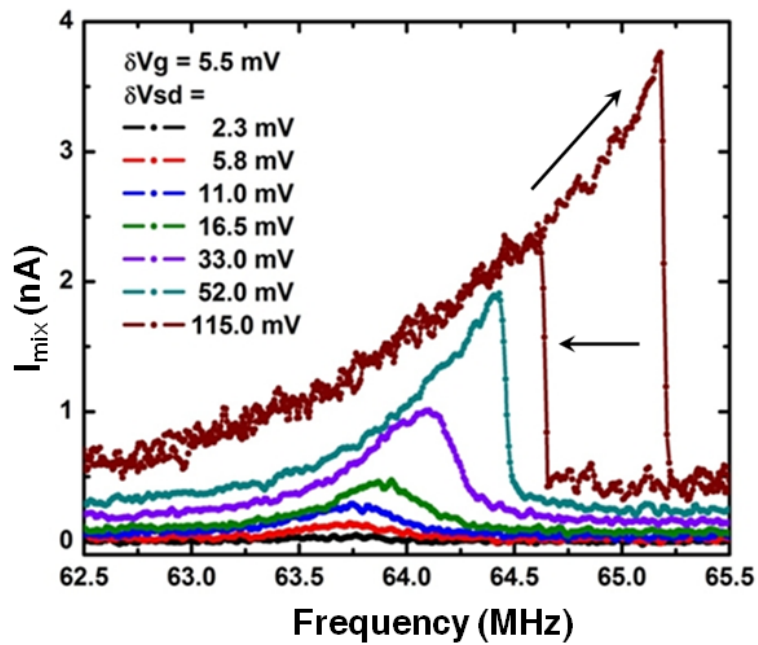

**Figure S3**. Nonlinear behavior with large drive power, bistable oscillation is observed when frequency is swept up and down. Mixing current is recorded in rms values.

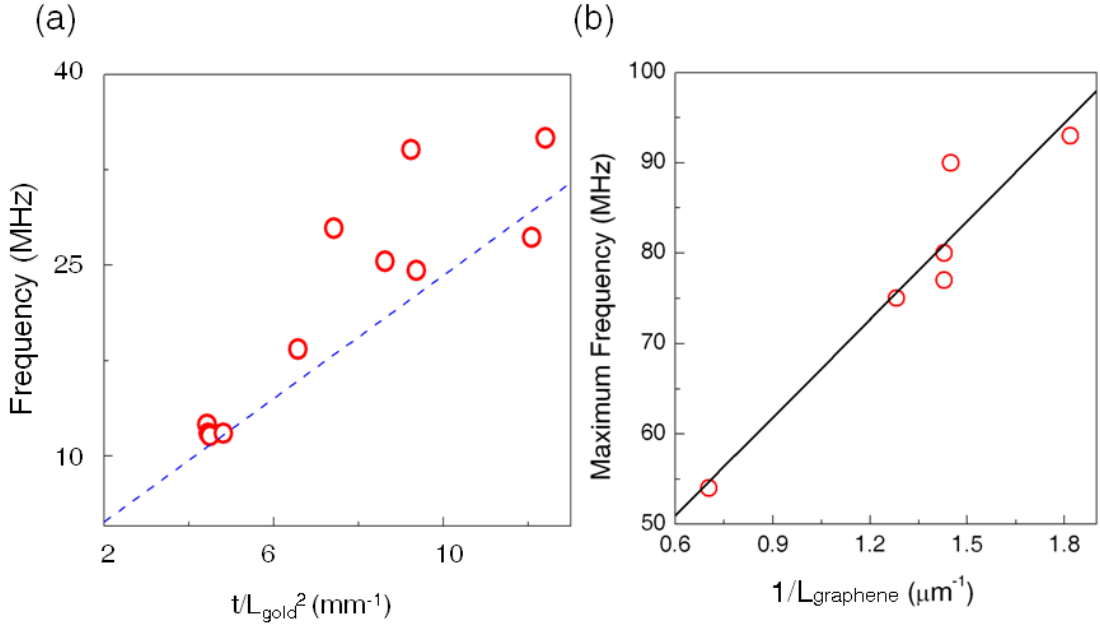

**Figure S4. a,** Observed gate-independent gold resonances, assigned to vibration of the gold electrodes, scale as $t/L_{gold}^2$, consistent with beam theory. The dotted line shows the expected value using $\sqrt{E/\rho} = 2400 m/s$ for gold. **b**, The maximum resonant frequency (at large $V_g$) scales inversely with length for devices made from same flake of graphene, consistent with a membrane model.

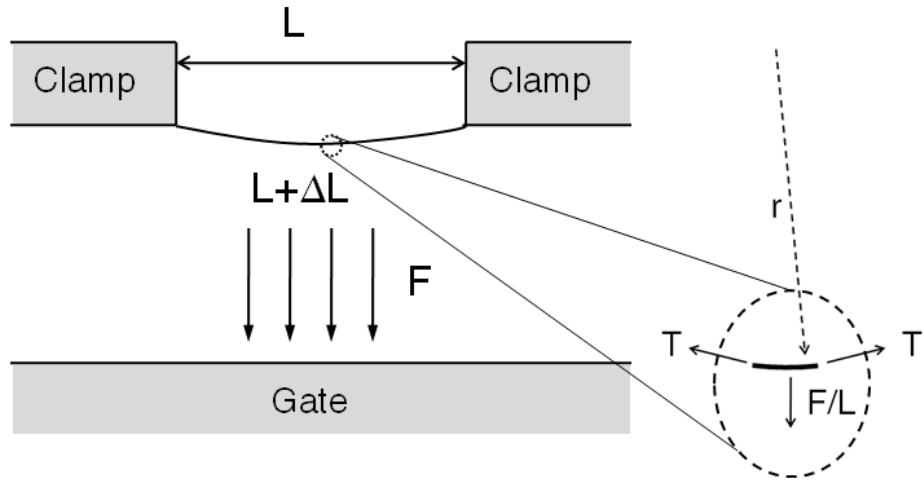

**Figure S5**. Continuum mechanics model for graphene resonator, simplified to 1D case. **L** is the length between the two metal clamps, **L+ΔL** is length of graphene under strain, **F** is the electrostatic force, **T** is the longitudinal tension, r is the radius of curvature.

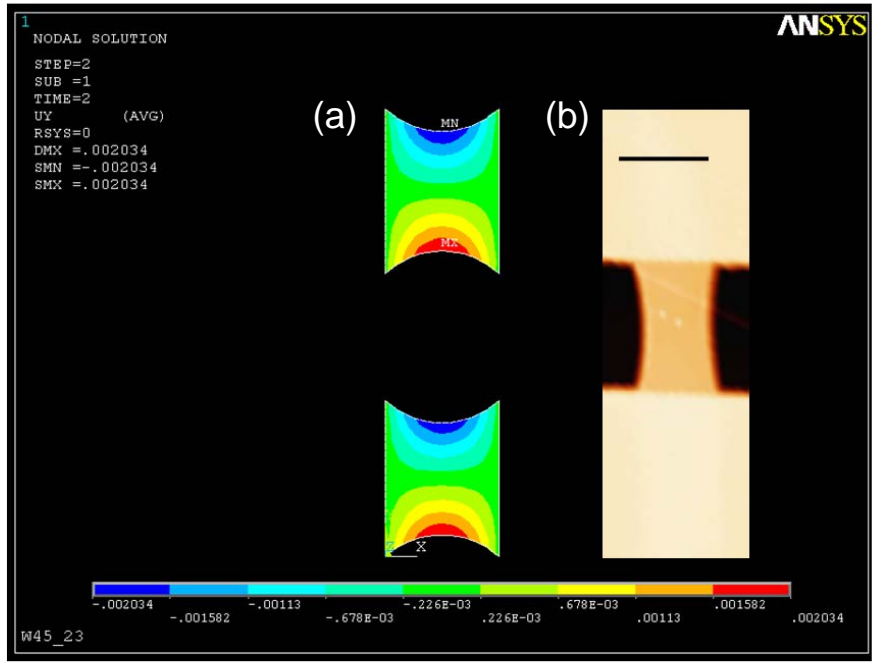

**Figure S6**. **a**, Deformation of the suspended metal contacts due to temperature change. $\Delta T = -100K$. **b**, AFM image of the same device at room temperature, scale bar: 1 μm.